\documentstyle[aps,amssymb,12pt]{revtex}
\begin{document}
\title{ Quantum and classical solutions for a \\
free particle in  wedge billiards}
\author{ A. G\'ongora-T$^{1,2}$,  Jorge V.
Jos\'{e}$^{2}$, S. Schaffner $^{2}$ and P. H. E.  Tiesinga$^{3}$}
\address{$^{1}$ Centro de Ciencias F\'\i sicas, Universidad Nacional
Aut\'onoma de M\'exico\\ Apartado Postal 48-3, 62250 Cuernavaca,
Morelos, MEXICO. \\ $^{2}$Physics Department and Center for Interdisciplinary
Research on Complex Systems, \\ Northeastern University, Boston,
MA~02115, USA.\\ $^{3}$Salk Institute, La Jolla, California, USA.
\\ }
\date{\today}
\preprint
\draft
\maketitle
\begin{abstract}
We have  studied the quantum and classical solutions of a particle
constrained to move inside a sector circular billiard with angle
$\theta_w$ and its pacman complement with angle $2\pi-\theta_w$.
In these billiards rotational invariance is broken and angular momentum
is no longer a conserved quantum number. The ``fractional" angular momentum
quantum solutions are given in terms of Bessel functions of fractional
order, with indices $\lambda_p={p\pi \over {\theta_w}}$, $p=1,2,...$ for the
sector and $\mu_q={q\pi \over {2\pi - \theta_w}}$, $q=1,2...$ for the
pacman. We derive a ``duality'' relation between both fractional indices
given by $\lambda_p={{p\mu_q} \over {2\mu_q - q}}$ and $\mu_q = {{q\lambda_p}
\over {2\lambda_p - p}}$.  We find that the average of the angular momentum
$\hat L_z$
is zero but the average of $\hat L^2_z$ has as  eigenvalues $\lambda_p^2$ and
 $\mu_q^2$.
We also make a connection of some classical solutions to their quantum
wave eigenfunction counterparts.

\end{abstract}
{Pacs 05.45.+b, 03.65.-w,72.20.Ht}
\newpage
Significant progress has been made in the last few years in
understanding the connection between classical and quantum solutions
for problems that show chaotic behavior.  A particularly important role
in this progress has been played by studies
in billiards with different geometries \cite{casati,bohigas,houches2,stock}.
The dynamics of a
free particle in a circular billiard is
completely integrable since energy and angular momentum are conserved.
A simple change of the circular billiard to the stadium billiard
immediately leads to
 chaotic particle dynamics \cite{stock}.
The analysis of these billiards eigenvalue spectra and eigenfunctions
have yielded a
number of clearly defined quantum manifestations of classically chaotic
Hamiltonians \cite{casati,bohigas,stock}. There are also other
types of billiards that,
although not being
explicitly chaotic,
can yield interesting novel quantum and classical behavior. A case in
point considered in this paper has a free particle that moves inside a boundary
defined by a wedge shaped section of a circular billiard. In this case
regular $2\pi$-rotational invariant
angular momentum is  no longer conserved. However, {\it fractional} quantum angular 
momentum
is well defined. It is this fractional angular momentum that makes this
problem interesting.  There have been other studies of modified circular
billiards, like a wedge sector in the presence of a constant
linear field that shows chaotic behavior \cite{wedge}, and circular chaotic billiards
with a straight cut \cite{linda}. These billiards are different from the
ones we study in this paper.
Here we consider sector billiards with angle $\theta_s=\theta_w$
and their pacman complement of angle $\theta_p=2\pi -\theta_w$.
These two types of
billiards  can be considered special cases of simple
wedge billiards that have corner discontinuities. For our
quantum analysis it is, however,  relevant to  separate them this
way  since the angular momentum spectral
properties of the  sector and its corresponding pacman
can be related to each other.

Our original motivation to study this problem actually came from experiments carried
out in pacman type mesoscopic billiards \cite{berry1,berry2}. These billiards
were studied in the presence of a magnetic field where the breaking of
rotational invariance precludes the full rotational closing of Aharonov-Bohm loops.
This significantly affects their contribution to the magnetoresistence.
As a first step towards understanding the magnetic field problem, we consider in this paper
the quantum and classical zero field cases. The quantum problem can be studied
to some extent analytically,
leading to interesting duality spectral relations. We also make a connection between the
quantum and classical solutions that help to understand the quantum results.

The quantum Hamiltonian for a free particle of mass $M$ in a  wedge billiard is
$\hat H = {\hat P^2 \over 2M} + V(r )$, with

\[
V(r)=\left\{
\begin{array}{ll}
0 & r\varepsilon D, \\
\infty  & r\notin D.
\end{array}
\right.
\]
Here $D$ is the domain of the wedge billiard.
The corresponding time-independent Schr\"{o}dinger equation is
\begin{equation}
(\nabla^2 + k^2)\Psi_{D} = 0,
\end{equation}
\noindent
with boundary condition $\Psi_{C}=\Psi(r\in C)= 0 $, with $C$  the
boundary of the domain $D$. Here $k^2={2ME \over {\hbar ^2}}$,
$E$ is the energy and
$2\pi \hbar =h$  is Planck's constant.
 We can immediately write the full set of solutions for the sector and
pacman wave-functions that satisfy the radial and angular boundary
conditions. The general normalized sector wave function is then given by
\begin{equation}
\Psi_{s}=\Sigma_{p}^{\infty} \Sigma_{n>1}^{\infty}
{J_{\lambda _p}({\alpha_{\lambda_p , n} \over r_w} \rho) \sin\lambda_p
\theta  \over  {\sqrt{\theta_s}\over 2}{r_w}
J_{\lambda_{p+1}}(\alpha_{\lambda_p,n})}.
\end{equation}
\noindent
Note that this wave function vanishes when $\theta =0$ and
$\theta=\theta_s$, and
$\rho={r_w}$, with $r_w$ the sector radius. The angular boundary
condition determines the values for the indices $\lambda_p$ that are
\begin{equation}
\lambda_p ={ p\pi \over \theta_s} ,\;  p=1,2,3...
\end{equation}
\noindent
Consider, for example,  the case
where the angular momenta are the integers
$\lambda_p=ap$ that correspond to the
 sector angles
$\theta_s={\pi/a}$,  with $a$ an integer.
Depending on the value of $a$ we will have a set of integers
that will be a subset of the index values
for the angular momenta
 of the full circle. For example, if we take
$\theta_s={\pi/4}$, we get $\lambda_p=4,8,12,16,...$ or if
$\theta_s={\pi/3}$ we get $\lambda_p=3,6,9,12,15,...$.
If the index is even we get a subset of even integer angular momenta
while for the odd case we get a subset of even and odd values for the
angular momentum. We can instead chose $\theta_s={b\over a}\pi$, which
will give $\lambda_p=p\frac{b}{a}$, with $b$ and $a$ prime numbers. In
this case $\lambda_p$ will generally be fractional and no full
circle angular momentum values will be present.
In the case where the angle is irrational, say
$\theta_s=\alpha\pi$, with $\alpha$ irrational,
the situation radically changes since $\lambda_p=p/\alpha$
has no corresponding analog in the angular momenta for the complete
circle or for the rational angles.

In the pacman case we can also write the complete wave function as
\begin{equation}
\Psi_{p}=\Sigma_{\mu_q}^{\infty} \Sigma_{m>1}^{\infty}
                 {J_{\mu_q}({\alpha_{\mu_q , m} \over {r_w}} \rho)
 \sin\mu_q \theta
                    \over
   {\sqrt{2\pi-\theta_w
}\over 2}{r_w} J_{\mu_{q+1}}(\alpha_{\mu_q , m})},
\end{equation}
\noindent
where $\mu_q$ is the order of the Bessel function and $m=1,2,3....$,
with $\theta_p=2\pi-\theta_w$, and
\begin{equation}
\mu_q={q\pi \over (2\pi - \theta_w)},\;  q=1,2,3...,
\end{equation}
which is of the same form
as in the sector case. Again as in that case this wave function
satisfies the imposed angular and radial boundary conditions. Note
that if we take the pacman angle equal to ${\pi/4}$,
 or equivalently as a ${7\pi/4}$ wedge, then
$\mu_q=4/7,8/7,12/7,16/7,...$ that  has even numerators
as in  the sector case but  with a fractional angular
momentum index that has no counterpart in the sector
nor in the circle cases. For an angle $\theta_w=\pi/a$, with $a$ an integer, we
have $\mu_q=q/(2a-1)$, $a\geq 1$. Or, more generally,
for $\theta_w={b\over a}\pi$,
with $b$ and $a$ prime numbers, we have $\mu_q={q\over (2a-b)}$, and
we need to have that $2a\not=b$, which is satisfied for prime numbers.
This is an interesting result since these fractional angular momentum
cases do not correspond to cases  previously studied in
group theory \cite{groups},
at least not to the best of our knowledge.
In the irrational case
with $\theta_w=2\pi-\alpha\pi$, we have $\mu_q=q/\alpha$, which has the
same value as in the sector case. In fact, we show from geometric relations between
the sector and the pacman wedges and for the same $\theta_w$,
the ``duality'' relations between the corresponding
Bessel function fractional angular momentum indices given by
\begin{equation}
\mu_q={\lambda_p \over {2\lambda_p - p}}q\Leftrightarrow \lambda_p =
      {\mu_q \over {2\mu_q - q}}p.
\end{equation}
\noindent
We note that, although the energy eigenvalues for the sector and the
pacman given by
$E_{\mu{_q}m}={\hbar \alpha^2_{\mu_q,m}\over 2M}$
and
$E_{\lambda{_p}n}={\hbar \alpha^2_{\lambda_p,n}\over 2M}$,
\noindent
are not the same, this duality relation gives a nontrivial
connection between fractional Bessel functions and the corresponding
``fractional'' angular momenta for the sector and its pacman complement.
In the wedge billiards rotational invariant
angular momentum is not a good quantum number. We calculate
then the average of the $z$-component of the angular momentum
$\hat L_z=(\frac{\hbar}{i}){\partial \over \partial \theta}$,
using the wave function given in Eq.(2), and we get
\begin{equation}
<\Psi_{s}\mid{\hat L_z}\mid \Psi_{s}>=
(\frac{\hbar}{i}){\sin^2\lambda_p\theta_w \over \theta_w}=0,
\end{equation}
and also for the pacman
\begin{equation}
<\Psi_{p}\mid{\hat L_z}\mid \Psi_{p}>=
(\frac{\hbar}{i}){\sin^2\mu_q(2\pi-\theta_w) \over (2\pi- \theta_w)}=0.
\end{equation}
\noindent
These results can be physically understood from a semi-classical
analysis. In a full circle we have that the particle  motion
completes a full rotation between $0$ and $2\pi$. For the wedges
the angular motion is limited to be between $\theta \in [0,\theta_w]$
for the sectors or $\theta \in [\theta_w,2\pi]$ for the
pacmen. Since in the sector the
particle moves from $0$ to $\theta_w$ and then bounces back to move from
$\theta_w$ to $0$,  the motion is librational rather than
rotational. 
One result of this is that for the separable eigenfunctions given above all have
zero average $L_z$. When the particle  bounces off the outer periphery, only $v_r$
changes and $L_z$ is unchanged. When it bounces off either wedge
wall, $v_\theta$ changes sign and thus  ${L_z}^2$ is unchanged.
This quantum result comes from our use of a specific coordinate system
with origin of rotation at the apex of the billiard. However, in the classical analysis
described below we also find internal closed orbits in the  billiards
where the angular momentum is not zero. To represent the latter ones
we need to write  a linear combination of the apex centered eigenfunctions
and carry out a coordinate transformation to define the new angular
momentum with center of rotation away from the apex.

Because of the $v_r$ cancellations with the boundary collisions
we are led to consider instead $\hat L_z^2$ that
would take care of these cancellations. Using the full sector
and pacman wave-functions we get
\begin{equation}
<\Psi_{s}\mid (\hat L_z)^2\mid \Psi_{s}>=
      -(\frac{\hbar}{i})^2 \lambda_p^2,
\end{equation}
\noindent
and
\begin{equation}
<\Psi_{p}\mid (\hat L_z)^2\mid \Psi_{p}>=
      -(\frac{\hbar}{i})^2 \mu_q^2.
\end{equation}
\noindent
We then see that the square of the Bessel function indices are good
quantum numbers. Note that this result applies even in the
irrational $\theta_w$ angle case.
Additionally, we also evaluated the expectation value of the quantum
mechanical currents defined by
$J\sim Im [(\Psi\nabla)^\ast\Psi]$. We found that the angular
component of the current
is zero, contrary to what happens in
the circle case, while the radial current component
is zero in
both billiard types, as in the circular case.

\noindent
We now make a qualitative connection of these quantum results to their classical counterparts.
Consider a wedge cut from a thin disk with radius $r_{w}$.  A particle
moves freely within the wedge, bouncing elastically from the walls.
Define a polar coordinate system $(\rho ,\theta)$
with origin at the apex of the wedge billiard and with $\theta=0$
pointing along the 
right hand boundary of the wedge. The particle is
constrained to move in the ranges $0\leq\theta\leq\theta _{w}$
and $0<\rho<r_{w}$. The classical motion of the particle is complicated in polar coordinates,
but the elastic collisions are simple to describe. Suppose the particle has velocity $(v_{\rho},v_{\theta})$ when it
encounters the maximum radius of the wedge. The outer circle normal
vector is perpendicular to $\hat \theta$,
so  $v_{\theta}$  will be unchanged. Because the collisions are elastic,
the energy ${m\over 2}(v_{\rho}^{2}+(r{v}_{\theta })^{2})$ must also be constant.
This requires that $v_{\rho}\Rightarrow-v_{\rho} $.  Similarly, for a collision
with a wedge radial wall,
$(v_{\rho},v_{\theta })\Rightarrow(v_{\rho},-v_{\theta}) $.
We can use dimensionless coordinates where $r_{w}=1$
 and $\left|\vec v\right|=1$, since the geometry of
 the trajectory is not changed by the scaling of time and radius.
In fact, the collisions with the radial walls can be removed by lifting
copies of the wedge onto an  infinite spiral in $\theta$.
The wedge copies are connected on the spiral in such a way that the motion of a particle
The spiral
is then projected onto the wedge by folding it like a fan.
The mapping from the spiral to the wedge is done in two stages. First,
define $\sigma\, \doteq\, \theta \,mod\, 2\theta_{w}$. Then

\[
\theta=\left\{
\begin{array}{ll}
 \sigma, & 0\leq \sigma \leq \theta _{w} \\
 2\theta _{w}-\sigma, &
\theta _{w}\leq \sigma <2\theta _{w}.
\end{array}
\right.
\]
In order for an orbit to close, the
particle must bounce off exactly the same point on the periphery,
going in the same direction.  This requires that the total angle
traversed must
be $b 2 \theta_w$, for some positive integer $b$.
Combining the two, a closed orbit is possible when $a \theta_c= b 2
\theta_w$. It
is easy to show that orbits will be closed or
open independently of the starting position of the first orbit. Orbits
can then
be characterized by the pair of positive integers
$(a,b)$ \cite{robinett}.  The chord angle $\theta_c$ is related to the angle of incidence
$\phi$ 
by $\theta_c =2\phi$.  In order for the chord to stay
within the wedge, $ \theta _c\leq \pi $,
with $n \geq {{ 2 \theta_w\over\pi}b}$. This implies, in particular, that the trajectory 
will be repeating if and only if $\theta_{w}$ and $\theta_{c}$ are commensurate.
The particle makes a series of collisions with the outer perimeter
at regular intervals $a \theta_c$.

We can now  compare the spectra of two wedges to see if they both have
orbits for certain common values of $\theta_c$.  In particular, consider a
wedge sector with angle $\theta_s$ and its $\theta_p = 2\pi-\theta_s$
complement pacman. In order for an orbit in each to share a common $\theta_c$,
 we need to have $
\theta _c={{b_s2\theta _s}
\over n_s}
={{b_{p}2\theta_p}
\over n_{p}}
$  or
$
{n_p\over b_p}
={\theta _p\over \theta _s}
\, {n_s\over b_s}
$. Thus, if the pacman complement angle is an integer multiple $m$ of the
angle of the sector, any orbit $(a,b)$ in the wedge
will have an orbit $(m*a,b)$ in the pacman for exactly the same value of
$\theta_c$.

In Fig. (\ref{fig1}) we show an example of a pacman rational angle case
with angle of $\pi/4$. In this figure we show the orbit $(16,3)$, where
the particle strikes the outer circle boundary 16 times before the orbit closes. During
such a traversal, the total rotation of the particle on the lifted spiral is equal
to $6\cdot \frac{7\pi}{4}$. We can also generate a whole family of classical
orbits for this pacman with indices like $(7,1)$, $(8,1)$, $(28,1)$, $(28,3)$,
 $(28,5)$ and so on.
In Fig. (\ref{fig2}) we show the corresponding eigenfunction for this pacman
case. Note that the caustic radius and the number of small and large
triangles correlate to the wave function amplitude densities shown in this
figure. The eigenfunction calculations were carried out by solving directly
the Shr$\ddot o$dinger equation using the finite element method.
In order to make a direct correlation between classical orbits and
quantum eigenfunctions we need to consider all the possible allowed
classical and quantum solutions for different values of the caustic
radius and set of parameter orbits $(a,b)$ and their corresponding quantum
counterparts. We do not carry out this analysis since we already know
the exact quantum solutions and we have described some of the classical
solutions to further understand their corresponding quantum
counterparts.

In conclusion, we have considered the quantum and classical problems of a 
free particle in  wedge billiards that exhibit fractional angular momentum due to
the breaking of rotational invariance. We found
 a new duality relation between the ``fractional'' indices for the
angular momentum Bessel functions for the sector and its complement
pacman billiard. We showed that $\hat L_z^2$
becomes a good quantum operator that can be used to characterize
the fractional angular momentum eigenvalues of these billiards.
We will treat elsewhere the classical and quantum dynamics of a charged 
particle in wedge billiards in the presence of a
constant homogeneous magnetic field. In that case there is a full transition from
integrable to  chaotic behavior as the magnetic field increases and new states
of chaotic whispering gallery modes appear in the large field limit.

\thanks

The work by AGT has been supported in part by CONACYT 3047P and UNAM sabbatical
grants, M\'exico. JVJ thanks NSF for partial financial support.
\newpage

\newpage

\vskip 1cm
\begin{figure}[hp]\begin{center}
\caption{
Here we show the orbit $(16,3)$ for the $\frac{\pi}{4}$ pacman
described in the text. The particle strikes the outer circle 16
times in 3 rotations to close the orbit completely. The total angle
covered in this orbit is equal to $6\cdot \frac{7\pi}{4}$.
\label{fig1}}
\end{center}\end{figure}

\begin{figure}[hp]\begin{center}
\caption{
In this figure we show the wave function density for the
$\frac{\pi}{4}$ pacman that geometrically corresponds to the classical
orbit of Fig.(1). Note that the number of small and large triangles, 
as well as the radius of the inner caustic,
agree with the ones shown in Fig. (1).
\label{fig2}}
\end{center}\end{figure}

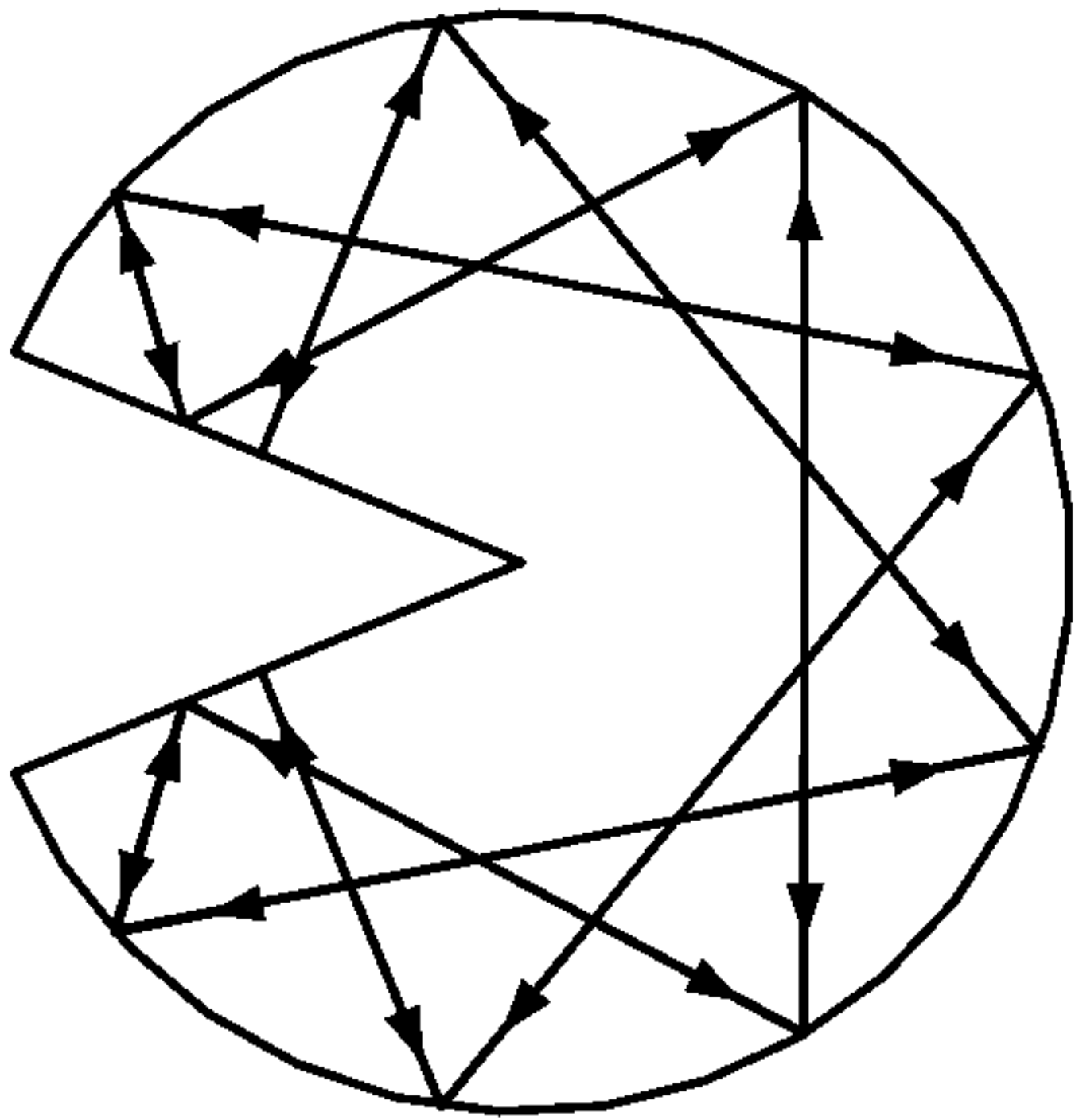
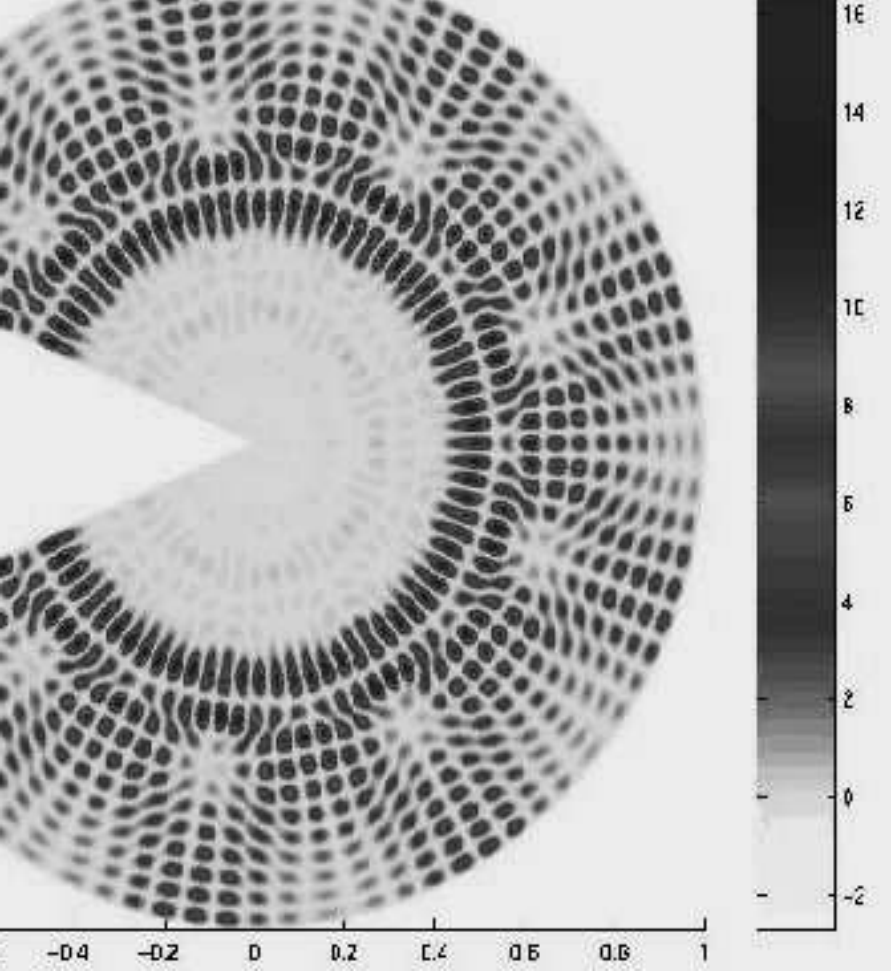
\end{document}